\documentclass[preprint]{elsarticle}

\usepackage{amsmath, amsfonts, amssymb, makeidx, graphicx, float, color, url, setspace, subfigure, rotating, lineno}
\usepackage[bookmarks, colorlinks, breaklinks]{hyperref}
\hypersetup{linkcolor = red, citecolor = blue, filecolor = black, urlcolor = blue}
\hypersetup{pdfauthor = {Chiranjit Mitra}}
\usepackage[section]{placeins}
\usepackage{epstopdf}

\modulolinenumbers[10]

\journal{Chaos, Solitons \& Fractals}

\begin{document}

\begin{frontmatter}

\title{Dynamical behaviors in time-delay systems with delayed feedback and digitized coupling}

\author{Chiranjit Mitra\fnref{myfootnote_1}}
\address{Indian Institute of Science Education and Research, Kolkata 741246, India}
\fntext[myfootnote_1]{chiranjitmitra4u@iiserkol.ac.in}

\author{G. Ambika\fnref{myfootnote_2}}
\address{Indian Institute of Science Education and Research, Pune 411008, India}
\fntext[myfootnote_2]{g.ambika@iiserpune.ac.in}

\author{Soumitro Banerjee\fnref{myfootnote_3}}
\address{Indian Institute of Science Education and Research, Kolkata, India - 741246, and King Abdulaziz University, Jeddah, Saudi Arabia}
\fntext[myfootnote_3]{soumitro@iiserkol.ac.in}

%----------------------------------------------------------------------------------------
%	ABSTRACT
%----------------------------------------------------------------------------------------

\begin{abstract}
We consider a network of delay dynamical systems connected in a ring via unidirectional positive feedback with constant delay in coupling. For the specific case of Mackey–Glass systems on the ring topology, we capture the phenomena of amplitude death, isochronous synchronization and phase-flip bifurcation as the relevant parameters are tuned. Using linear stability analysis and Master Stability Function approach, we predict the region of amplitude death and synchronized states respectively in the parameter space and study the nature of transitions between the different states. For a large number of systems in the same dynamical configuration, we observe splay states, mixed splay states and phase locked clusters. We extend the study to the case of digitized coupling and observe that these emergent states still persist. However, the sampling and quantization reduce the regions of amplitude death and induce phase-flip bifurcation.
\end{abstract}


\end{frontmatter}

\linenumbers

%----------------------------------------------------------------------------------------
%	INTRODUCTION
%----------------------------------------------------------------------------------------

\section{Introduction}

Dynamical systems with delay due to finite signal transmission times, switching speeds and memory effects are ubiquitous in nature. In recent years, such delay systems have attracted the attention of various domains of science and engineering such as physics, electronics, biology, ecology, economics etc \cite{TDS_Lakshmanan}.

Time-delays arise frequently in the interaction between the systems of a network that model complex systems. The inherent delay in dynamical systems as well as delays in the interaction of such systems of a network makes analysis of networks of delay coupled delay dynamical systems a subject of significant interest. Moreover, the analysis of time-delayed dynamics on networks holds immense potential in understanding a large class of real-world systems from science and engineering such as coupled laser arrays \cite{TDS_Lasers}, gene regulatory networks \cite{TDS_Gene_Regulatory_Networks}, neuronal networks \cite{TDS_Neuronal_Networks}, complex ecosystems \cite{TDS_Ecology_Networks} etc. However, the mathematical analysis of delay dynamical systems is challenging owing to the fact that even a single delay differential equation is representative of a theoretically infinite (numerically high) dimensional system. Further, the complexity associated with the network structure makes the analysis formidable, leaving this area less explored. Recently, there has been a drive towards exploring various emergent phenomena such as synchronization \cite{Pikovsky, Synchronization_Networks_Review}, amplitude death \cite{AD_Review, AD_Konishi_Delay_Ring, Networks_DCDO}, phase-flip bifurcation \cite{PF_Prasad, PF_Universal, PF_Neurons}, splay states \cite{Splay_State} and cluster formation \cite{Clustering} displayed by a collection of coupled delay dynamical systems.

Previously, the effect of gradient coupling on oscillation death in ring networks of delay coupled oscillators has been investigated in \cite{AD_Gradient_Coupling}. Further, \cite{AD_Processing_Delay} discusses the effect of a processing delay in the coupling, which can inhibit AD in a network of coupled oscillators. In sharp contrast to propagation delay which has the tendency to induce AD, processing delay revives the oscillations in the AD regime to allow for sustained rhythmic functioning of a network. Further, \cite{Splay_State_Local_Global_Coupling} focusses on the splay states in a ring of coupled nonlinear oscillators for different cases of local (nearest-neighbor) coupling to global (all-to-all) coupling.

In this paper, we demonstrate the dynamics offered by a ring network of nonlinear delay dynamical systems, like synchronization, amplitude death (AD), phase-flip, splay states and phase locked clusters. We consider the interacting systems coupled via unidirectional positive feedback connections with constant delay in coupling.

Among the various regular network topologies studied, ring topology forms the basic structure in many complex real-world systems from natural science and engineering like single crystal nanorings with semiconductor and piezoelectric characters \cite{Ring_1}, Central Pattern Generators in locomotion, peripheral neural systems etc \cite{Rings_biological, Rings_neural}. Most of the studies in coupled systems are with diffusive coupling between systems. In this work, we consider positive feedback as the interaction connecting systems. Positive feedback loops, identified as a self-reinforcing chain of cause and effect find applications in electronics in regenerative circuits \cite{F_Circuits}, generation of nerve signals and blood clotting in physiology \cite{F_Physiology}, several processes in evolutionary biology \cite{F_Evolution}, system risk in finance, vicious and virtuous circles \cite{F_V_V_Cycle}, climate system and carbon cycle in climatology \cite{F_Climatology} etc. This gives sufficient motivation for studying the dynamics offered by positive feedback loops in ring networks of delay coupled delay dynamical systems.

The analysis reported here illustrates the occurrence of a host of interesting phenomena such as AD, complete and phase synchronized states, phase-flip, splay states and clustering. In a practical implementation of such networks, dynamical systems are synchronized using sampled data controllers which only require the samples of the state variables at discrete (sampled) time instants. It is known that the sampling time significantly impacts the stability and performance of an analog to digital converter, price of implementing it in terms of CPU-time consumption etc \cite{Hybrid_chaos_synchronization}. Given that chaotic systems are highly sensitive to sampling, it further raises the importance of investigating the digital implementation of analog signals. Subsequently, there has been a recent drive towards exploring the synchronization of such hybrid systems consisting of continuous dynamical subsystems interacting via digital connections \cite{Sporadic_driving_1, Sporadic_driving_2, Sporadic_driving_3, Digital_transmission_chaos_generators, Lure_sampled_control, Lure_delayed_sampled_control, Synchronization_CDNs}.

We study a hybrid version of our model where the signal from one node is digitized (sampled and quantized) before another node receives it. We explore the dynamics offered by the hybrid model in the presence of coupling delay and study the effect of changing the sampling time and step size of quantization of the discrete signal being transmitted. To the best of our knowledge, we give the first report on the occurrence of phase-flip bifurcation on account of changing the sampling time and step size of quantization.

The paper is organized as follows: In section \ref{sec:Model}, we describe the basic model. In section \ref{sec:ED}, we report the emergent dynamics exhibited by the system. In section \ref{sec:NT}, we study the nature of transitions between the emergent dynamics observed in the previous section. In section \ref{sec:PLC}, we report the emergent dynamics observed for a large number of systems on the network.
In section \ref{sec:Model_Hybrid}, we report and investigate the emergent dynamics observed with the digitization of coupling between the nodes of the network. Finally, we summarize the relevant observations in section \ref{sec:Conclusion}.

%----------------------------------------------------------------------------------------
%	MODEL
%----------------------------------------------------------------------------------------

\section{Model} \label{sec:Model}

We consider a network of $N$ Mackey-Glass (MG) systems where the evolution of each isolated unit is given by:
\begin{equation} \label{eq:MG_1}
\dot{x} = \frac{ax_{\tau_s}}{1+x_{\tau_s}^{c}} - bx(t);\ x_{\tau_s} = x(t-\tau_s)
\end{equation}
where $a$, $b$ and $c$ are positive constants and $\tau_s$ is the positive constant inherent time-delay in individual systems \cite{MG_system}. This system was introduced as a model for blood production in patients suffering from Leukaemia where $x(t)$ represents the concentration of blood (density of mature cells in bloodstreams) at time $t$ and the delay time $\tau_s$ is the interval between the maturation of red blood cells (RBCs) for release in bloodstreams after their production in the bone marrow. It is a paradigm of delay dynamical systems and has been extensively studied in the literature for its chaotic and hyperchaotic behaviour \cite{TDS_Lakshmanan}. Its study has been experimentally realized using analog electronic circuits \cite{MG_circuit_1, MG_circuit_2}.

\begin{figure} [!htbp]
\begin{center}
\includegraphics[scale=0.5]{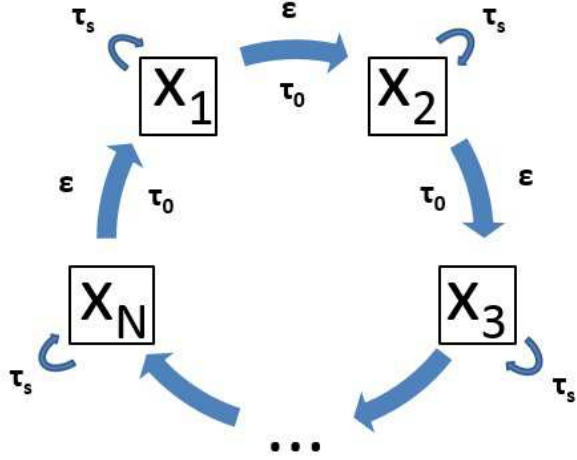}
\caption{\label{fig:Model}(Colour online) Network of $N$ MG systems each represented by the dynamical variable $x_i$ where $i = 1,\ 2, \ ...\,\ N$ and constant inherent delay $\tau_s$ connected in a ring with coupling strength $\epsilon$ such that the coupling simply provides a positive feedback from one system to the successive system on the ring with a delay $\tau_0$.}
\end{center}
\end{figure}

We consider a network consisting of $N$ MG systems each with constant inherent delay $\tau_s$ connected in a ring with coupling strength $\epsilon$ such that the coupling simply provides a unidirectional positive feedback from one system to the successive system on the ring with a delay $\tau_0$. This is shown schematically in figure \ref{fig:Model}.
%%%%%%%%%%%%%%%%%%%%
%%%%%%%%%%%%%%%%%%%%
Consequently, the dynamical equation of the network is given by:
\begin{equation} \label{eq:Model}
\dot{x}_i = af(x_{i\tau_s})-bx_i + \epsilon x_{j\tau_0}
\end{equation}
where
\begin{equation} \label{eq:MG_2}
 f(x_{\tau_s}) = \frac{x_{\tau_s}}{1+x_{\tau_s}^{c}} 
\end{equation}
for $i = 1, 2, 3, \ldots, N-1, N\ \&\ j = N, 1, 2, \ldots, N-1$ respectively. We explore the possibility of occurrence of various emergent phenomena in this system by stability analysis and direct numerical simulations.

%----------------------------------------------------------------------------------------
%	EMERGENT DYNAMICS
%----------------------------------------------------------------------------------------

\section{Emergent dynamics in the model} \label{sec:ED}

We start with the simplest case of $N = 2$ where the model of figure \ref{fig:Model} reduces to the case of 2 systems connected to each other via bidirectional delayed positive feedback. By direct numerical analysis of this system, we find that depending upon the coupling strength ($\epsilon$) and time delay in coupling ($\tau_0$) the system is capable of exhibiting complete synchronization, phase-flip and AD.

\begin{figure} [!htbp]
\begin{center}
\includegraphics[scale=0.2]{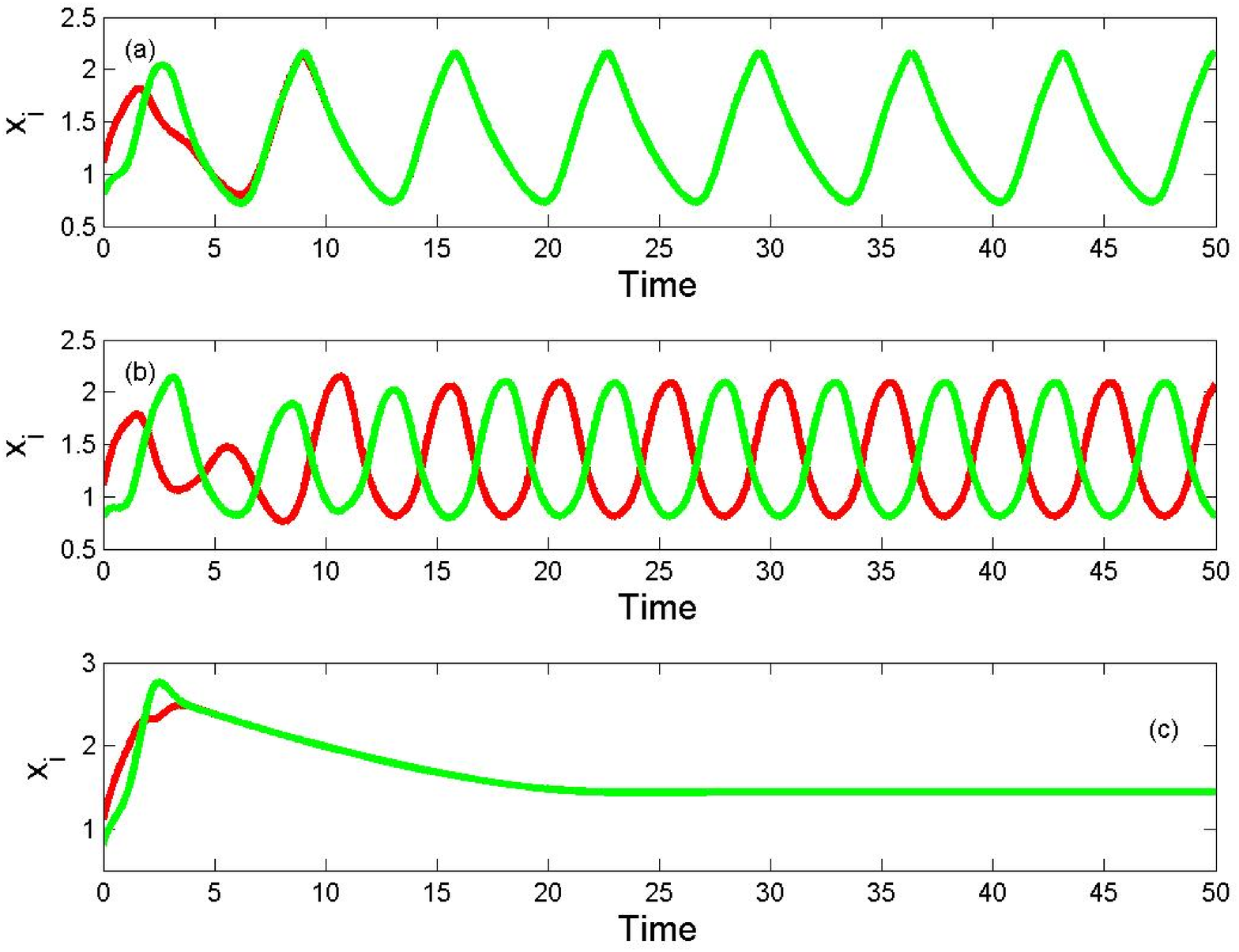}
\caption{\label{fig:MG_eq_net_modu_2_eps_tau0_Chaotic_TS}(Colour online) Time series for 2 systems exhibiting (a) complete synchrony at $\epsilon = 0.6$, $\tau_0 = 0.4$ (b) phase-flip at $\epsilon = 0.6$, $\tau_0 = 1.2$ (c) AD at $\epsilon = 0.95$, $\tau_0 = 0.4$. Red and green colors represent systems 1 and 2 respectively.} 
\end{center}
\end{figure}

The parameter values of the MG system used for carrying out the numerical simulations are: $a = 2.0$, $b = 1.0$ and $c = 10.0$. For these parameter values, a MG system has a fixed point attractor for $\tau_s < 0.471$, limit cycle for $0.471 < \tau_s < 1.33$, period doubling sequence for $1.33 < \tau_s < 1.68$ and chaos for $\tau_s > 1.68$ \cite{MG_system_parameters}. We keep the value of $\tau_s = 2.0$ such that individual MG systems are in the chaotic regime. The computational scheme used here for numerically integrating our model in (\ref{eq:Model}) is described in \cite{Gyori_delay_paper}. The time series thus obtained for certain parameter values of ($\epsilon,\ \tau_0$) corresponding to different emergent dynamics is shown in figure \ref{fig:MG_eq_net_modu_2_eps_tau0_Chaotic_TS}.

The state of complete synchronization between two time series $x(t)$ and $y(t)$ is measured by the cross correlation (CC) function defined as:
\begin{equation}
CC = \frac{ <(x(t)-<x(t)>) (y(t)-<y(t)>)> }{ \sqrt{ <(x(t)-<x(t)>)^2> <(y(t)-<y(t)>)^2> } }
\end{equation}
CC is equal to 1 in the completely synchronized state \cite{Pikovsky}. AD is characterized by using an index $A^{'}$ defined as difference between the global maximum and global minimum of the time series over a sufficiently long interval of time after neglecting the initial transients. The states characterized by $A^{'} = 0$ correspond to AD \cite{AD_measure}. The phases of individual systems has to be defined in order to precisely identify the phase synchronized states. The instantaneous phase of a system is defined as \cite{Phase_measurement}:
\begin{equation} \label{eq:Phase_measurement}
\phi(t) = 2\pi \left( \frac{t-\tau_k}{\tau_{k+1}-\tau_k} \right) + 2\pi k;\  \tau_k \le t < \tau_{k+1}
\end{equation}
$\tau_k$ corresponding to the instant when the time series crosses some threshold level in one direction or attains a local maxima, are stored and a phase increase of $2\pi$ is attributed to each such crossing. The phase $\phi(t)$ and the phase difference between two systems $\psi(t)$ are calculated using (\ref{eq:Phase_measurement}). Then the mean phase difference $<\psi(t)>$ over many such cycles is calculated. The states characterized by $<\psi(t)> \sim 0$ correspond to in-phase synchronization while states with $<\psi(t)> \sim \pi$ and $<\psi(t)> \sim \frac{2\pi}{N}$ correspond to phase-flipped states for 2 and splay states for $N (> 2)$ systems respectively. As clear from the time series in figure \ref{fig:MG_eq_net_modu_2_eps_tau0_Chaotic_TS}, synchronization is achieved with control to amplify the periodic states.
%%%%%%%%%%%%%%%%%%%%
%%%%%%%%%%%%%%%%%%%%

%----------------------------------------------------------------------------------------
%	STABILITY ANALYSIS
%----------------------------------------------------------------------------------------

\subsection*{Stability analysis: amplitude death} \label{sec:Stability_Analysis}

In this subsection, using the tool of linear stability analysis we analyze and predict, the parameter values for which the coupled system of (\ref{eq:Model}) exhibits AD. Let $x^* = \left(x_1^*, x_2^*, ..., x_N^* \right)^T$ be an equilibrium point of the system where $x^*$ satisfies
\begin{equation}
bx_i^* = af(x_i^*) + \epsilon x_j^*
\end{equation}
for $i = 1, 2, 3, \ldots, N-1, N$ and $j = N, 1, 2, \ldots, N-1$ respectively. The equilibrium points of our system obtained by substituting $x^*_i = x^*_{i\tau_s} = x^*_{i\tau_0} = x^*\ \textrm{for}\ i=1, 2, ..., N$ (considering the symmetry in our model) in (\ref{eq:Model}) are:
\begin{equation} \label{eq:MG_FP}
x^* = 0,\ \left(\frac{a}{b-\epsilon}-1\right)^\frac{1}{c} 
\end{equation} 
The numerical results suggest that the equilibrium point $x^* = \left(\frac{a}{b-\epsilon}-1\right)^\frac{1}{c}$ is stabilized leading to AD. Henceforth, the stability of the equilibrium point $x^* = \left(\frac{a}{b-\epsilon}-1\right)^\frac{1}{c}$ is analyzed.

Locally linearizing (\ref{eq:Model}) around an equilibrium point $x^*$ by expanding it in a Taylor series about the same we obtain the following transcendental characteristic equation:
\begin{equation} \label{eq:CE}
\textrm{det} \left( \left( \lambda + b - af^{'}(x^*)  e^{-\lambda \tau_s} \right) I_N -\epsilon e^{-\lambda \tau_0}C \right) = 0
\end{equation}
where $f^{'}(x^*) = \left.\frac{df(x_{\tau_s})}{dx_{\tau_s}}\right\arrowvert_{x_{\tau_s}=x^*}$. 
Thus, for MG system of (\ref{eq:MG_2}), 
\begin{equation}
f^{'}(x^{*}) = \frac{(b-\epsilon)(a-c(a-(b-\epsilon)))}{a^2}
\end{equation}
for $x^* = \left(\frac{a}{b-\epsilon}-1\right)^\frac{1}{c}$. $I_N$ denotes an $N \times N$ identity matrix and $C$ called the \emph{adjacency matrix} essentially captures the topology of the network such that $C_{ij}$ is 1 if there is a link from $j$ to $i$ and 0 otherwise \cite{Networks_DCDO}. For the ring network under consideration: 
\begin{equation} \label{eq:Adjacency_matrix}
C = \left(\begin{array}{ccccc}
0 & 0 & \ldots & 0 & 1 \\
1 & 0 & \ldots & 0 & 0 \\
  & \ddots & \ddots & 0 & 0 \\
0 & 0 & \ddots & \ddots & \\
0 & 0 & \ldots & 1 & 0
\end{array}\right)_{N \times N}
\end{equation}
We further simplify the characteristic equation (\ref{eq:CE}) to obtain it as a product of $N$ terms. Let $\chi = e^{\iota\frac{2\pi}{N}}$, $v_j = \left(1, \chi^{j}, \chi^{2j}, \ldots, \chi^{(N-1)j} \right)^T$ where $j = 0, 1, 2, ..., N-1$ \cite{AD_SA}. We have 
\begin{eqnarray*}
\chi^{Nj} & = & 1 \\
\chi^{(N-1)j} & = & \chi^{-j} \\
\chi^{(N-2)j} & = & \chi^{-2j} \\
& \vdots &
\end{eqnarray*}
Thus,
\begin{equation}
C v_j = \left(\begin{array}{ccccc}
\chi^{(N-1)j} \\
1 \\
\chi^{j} \\
\vdots \\
\chi^{(N-2)j}
\end{array}\right)v_j
= \chi^{-j} v_j
\end{equation}
Hence,
\begin{equation}
\begin{split}
& \left( \left( \lambda + b - af^{'}(x^*)  e^{-\lambda \tau_s} \right) I -\epsilon e^{-\lambda \tau_0}C \right) v_j \\
& = \left( \left( \lambda + b - af^{'}(x^*)  e^{-\lambda \tau_s} \right) - \epsilon e^{-\lambda \tau_0}\chi^-j \right) v_j
\end{split}
\end{equation}
Therefore, the characteristic equation (\ref{eq:CE}) takes the form:
\begin{equation}
\begin{split}
& det\left( \left( \lambda + b - af^{'}(x^*) e^{-\lambda \tau_s} \right) I_N -\epsilon e^{-\lambda \tau_0}C \right) \\
& = \prod \limits_{j=0}^{N-1} \left( \lambda + b - af^{'}(x^*)  e^{-\lambda \tau_s} - \epsilon e^{-\lambda \tau_0}\chi^-j \right) \\
& = \prod \limits_{j=0}^{N-1} \Delta_{j}(\lambda) \\
& = 0
\end{split}
\end{equation}
where
\begin{equation}
\Delta_{j}(\lambda) = \left( \lambda + b - af^{'}(x^*)  e^{-\lambda \tau_s} - \epsilon e^{-\lambda \tau_0}\chi^-j \right)
\end{equation}
Substituting $\lambda = \nu + \iota \omega$; $\nu,\ \omega \in \Re,$ yields
\begin{equation}
\begin{split}
\Delta_{j}(\lambda) = & \Delta_{j}(\nu + \iota \omega) \\
= & (\nu + \iota \omega) + b - af^{'}(x^*)  e^{-(\nu + \iota \omega) \tau_s} \\
  & - \epsilon e^{-(\nu + \iota \omega) \tau_0}\chi^-j
\end{split}
\end{equation}
Separating real and imaginary parts,
\begin{equation}
\begin{split}
& \nu = -b + af^{'}(x^*) e^{-\nu \tau_s} \cos(\omega \tau_s) \\
+ &\epsilon e^{-\nu \tau_0} \left( \cos(\omega \tau_0) \cos \left( \frac{2 \pi}{N}j \right) - \sin(\omega \tau_0) \sin\left( \frac{2 \pi}{N}j \right) \right) \\
& \omega = -af^{'}(x^*) e^{-\nu \tau_s} \sin(\omega \tau_s) \\ 
- & \epsilon e^{-\nu \tau_0} \left( \cos(\omega \tau_0) \sin \left( \frac{2 \pi}{N}j \right) + \sin(\omega \tau_0) \cos \left(\frac{2 \pi}{N}j \right) \right)
\end{split}
\end{equation}
Simplifying, we get
\begin{equation} \label{eq:CE_nu_om}
\begin{split}
\nu = & -b + af^{'}(x^*) e^{-\nu \tau_s} \cos(\omega \tau_s) \\
& + \epsilon e^{-\nu \tau_0} \cos  \left( \omega \tau_0 + \frac{2 \pi}{N}j \right) \\
\omega = & -af^{'}(x^*) e^{-\nu \tau_s} \sin(\omega \tau_s) \\ 
& - \epsilon e^{-\nu \tau_0}  \sin \left(\omega \tau_0 + \frac{2 \pi}{N}j \right)
\end{split}
\end{equation}
The equilibrium point $x(t) = x^*$ of (\ref{eq:Model}) will be asymptotically stable if all roots of the characteristic equation (\ref{eq:CE}) have negative real parts. The equilibrium point loses stability only when one of the roots of the characteristic equation crosses the imaginary axis as the system parameters are varied. Thus, stability of the system changes in parameter space where the roots of (\ref{eq:CE}) have zero real parts. Thus, substituting $\nu = 0$ in (\ref{eq:CE_nu_om}) yields
\begin{equation} \label{eq:CE_nu_0}
\begin{split}
0 & = -b + af^{'}(x^*) \cos(\omega \tau_s) + \epsilon \cos  \left( \omega \tau_0 + \frac{2 \pi}{N}j \right) \\
\omega & = -af^{'}(x^*) \sin(\omega \tau_s) - \epsilon  \sin \left(\omega \tau_0 + \frac{2 \pi}{N}j \right)
\end{split}
\end{equation}
(\ref{eq:CE_nu_0}) can be numerically solved to obtain curve(s) in the $(\epsilon,\ \tau_0)$ parameter plane which define the boundary where the stability of the equilibrium point changes and onset/offset of AD occurs.

The sign of $Re \left( \frac{d\lambda}{d\epsilon} \right)$ at $\lambda = \iota \omega$ gives insight into the direction of crossing of the roots of (\ref{eq:CE_nu_0}) as $\epsilon$ is varied. The roots cross from left to right (right to left) in the Argand plane if the sign is positive (negative).
\begin{equation} \label{eq:dlambdadepsilon}
\begin{split}
& Re \left. \left( \frac{d\lambda}{d \epsilon}\right) \right\arrowvert_{\lambda = \iota \omega} \\
& = \frac{a \frac{df^{'}(x^*)}{d\epsilon} \cos \left( \omega \tau_s \right) + \cos \left( \omega \tau_0 + \frac{2\pi j}{N} \right)}{1 + af^{'}(x^*)\tau_s \cos(\omega \tau_s) + \epsilon \tau_0 \cos \left( \omega \tau_0 + \frac{2\pi j}{N} \right)}
\end{split}
\end{equation}
where
\begin{equation}
\frac{df^{'}(x^*)}{d\epsilon} = \frac{ac-a-2c(b-\epsilon)}{a^2}
\end{equation} 
Substituting in (\ref{eq:dlambdadepsilon}), the value of $(\epsilon$, $\tau_0)$ obtained by numerically solving (\ref{eq:CE_nu_0}) for the case of $N = 2$ systems investigated here we find that,
\begin{equation} \label{eq:Sign_dl_de}
\left. Re \left( \frac{d\lambda}{d \epsilon}\right) \right\arrowvert_{\lambda = \iota \omega} < 0
\end{equation}
implying that the roots cross from right to left in the Argand plane as $\epsilon$ is increased while crossing the boundary  of the AD region in the $(\epsilon,\ \tau_0)$ parameter plane.

%----------------------------------------------------------------------------------------
%	STABILITY ANALYSIS: SYNCHRONIZED STATES
%----------------------------------------------------------------------------------------

\subsection*{Stability analysis: synchronized states} \label{sec:MSF}

The dynamical equation of the network is given by (\ref{eq:Model}):
\begin{equation*}
\dot{x}_i = af(x_{i\tau_s})-bx_i + \epsilon x_{j\tau_0}
\end{equation*}
where
\begin{equation*}
f(x_{\tau_s}) = \frac{x_{\tau_s}}{1+x_{\tau_s}^{c}} 
\end{equation*}
for $i = 1, 2, 3, \ldots, N-1, N\ \&\ j = N, 1, 2, \ldots, N-1$ respectively. 

For the case of $N = 2$ systems, linearizing (\ref{eq:Model}) about a given synchronized state $\mathbf{\tilde{x}_s}$ we get:
\begin{equation} \label{eq:Linearized}
\Delta \dot{\mathbf{x}} = af^{'} (\mathbf{\tilde{x}_s}) \Delta \mathbf{x}_{\tau_s}  -b \Delta \mathbf{x} + \epsilon C \Delta \mathbf{x}_{\tau_0}
\end{equation}
where 
\begin{equation*}
\mathbf{x(t)} = \left(\begin{array}{ccccc}
x_{1}(t) \\
x_{2}(t)
\end{array}\right)
\end{equation*}
and $C$, the adjacency matrix in (\ref{eq:Adjacency_matrix}) becomes:
\begin{equation}
C = \left(\begin{array}{ccccc}
0 & 1 \\
1 & 0
\end{array}\right)_{2 \times 2}
\end{equation}

Now, diagonalizing the dynamical equation of the network in (\ref{eq:Linearized}) and using the ansatz $\Delta x_i = c_i v(t)$:
\begin{equation} \label{eq:MSF}
\dot{v} = af^{'} (\mathbf{\tilde{x}_s}) v_{\tau_s} -b v + \epsilon \lambda_m v_{\tau_0}
\end{equation}
where $\lambda_m$ is an eigenvalue of the adjacency matrix ($C$).

We need to compute the Lyapunov exponents of (\ref{eq:MSF}) in order to evaluate stability. The largest Lyapunov exponent (LLE) for a particular value of $\tau_0$ is called the \emph{Master Stability Function} (MSF). Although (\ref{eq:MSF}) depends on node index via $\mathbf{\tilde{x}_s}$, yet the synchronous solutions {$x_i$} are identical, apart from a temporal shift. Therefore, the LLE which is an asymptotically defined quantity is independent of the choice of the node index.

Each synchronized state $S_k$ corresponds to a particular relative phase ($\delta_k$) between successive oscillators e.g., $\delta_0 = 0$ for isochronally synchronized state ($S_0$), $\delta_1 = \pi$ for phase-flipped state ($S_1$), $\delta_2 = \frac{2 \pi}{N}$ for splay synchronized state ($S_2$) with $N (> 2)$ systems which have been observed in our model.

For a particular synchronized state $S_k$ to be stable, perturbations orthogonal to $l_k$ (eigenvector corresponding to the synchronized state $S_k$) must decay \cite{Isochronal_Splay_Experiment}. Therefore, we evaluate (\ref{eq:MSF}) for all $m = 1, 2, 3, \ldots, N-1, N$ except $m = k$ because we are interested only in perturbations orthogonal to the eigenvector corresponding to the synchronized state $S_k$. Thus, for $N = 2$, the eigenvector corresponding to the isochronally synchronized state ($S_0$) is   $l_0 = \left(\begin{array}{c}
1\\
1
\end{array}\right)$, so we ignore the Lyapunov exponent corresponding to the eigenvalue $\lambda_0 = 1$ when calculating the
MSF for the isochronally synchronized state. Similarly, the eigenvector corresponding to the phase-flipped state ($S_1$) is   $l_1 = \left(\begin{array}{c}
1\\
-1
\end{array}\right)$, so the Lyapunov exponent corresponding to the eigenvalue $\lambda_1 = -1$ is ignored when calculating the
MSF for the phase-flipped state. The LLE of (\ref{eq:MSF}) i.e., the MSF if less than zero, implies that the synchronized state under consideration is stable.

\begin{figure} [!htbp]
\begin{center}
\includegraphics[scale=0.2]{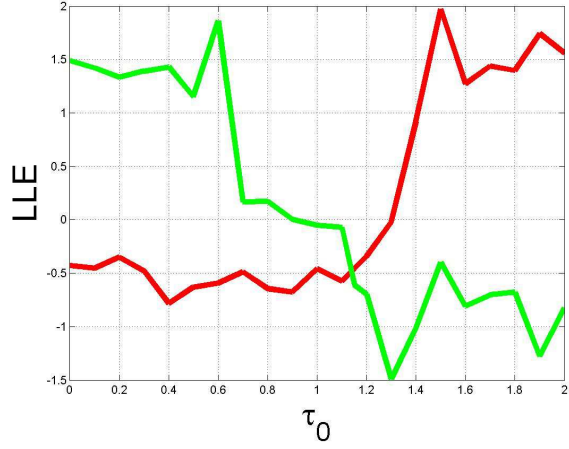}
\caption{\label{fig:MSF}(Colour online) Largest Lyapunov Exponent (i.e, Master Stability Function) of (\ref{eq:MSF}) as a function of coupling delay ($\tau_0$) for 2 systems at $\epsilon = 0.6$. The red and green lines represent completely synchronized and phase-flipped states respectively.}
\end{center}
\end{figure}

Figure \ref{fig:MSF} shows the LLE (MSF) of (\ref{eq:MSF}) as a function of coupling delay ($\tau_0$) for the simplest case of $N = 2$ systems at $\epsilon = 0.6$.
%%%%%%%%%%%%%%%%%%%%
%%%%%%%%%%%%%%%%%%%%

For smaller values of $\tau_0$ (away from the phase-flip region), the completely synchronized state is stable while for larger values of $\tau_0$ (away from the phase-flip region), the phase-flipped state is stable. However, the transition region from completely synchronized state to phase-flipped state has a window of  multi-stability where  both the states can be stable with different basins.

\begin{figure} [!htbp]
\begin{center}
\includegraphics[scale=0.2]{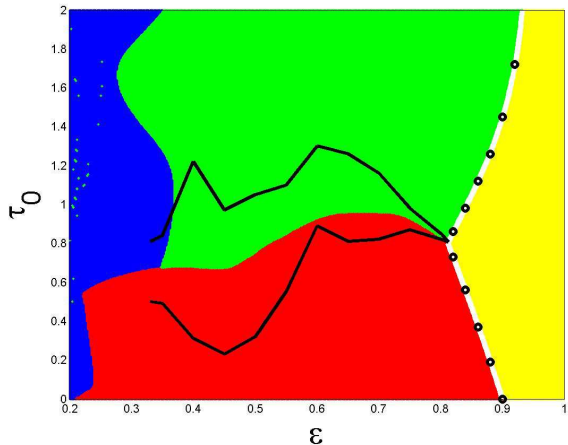}
\caption{\label{fig:MG_eq_net_modu_2_eps_tau0_Chaotic}(Colour online) Parameter plane of coupling delay ($\tau_0$) versus coupling strength ($\epsilon$) for 2 systems. The blue, red, green and yellow colored regions represent unsynchronized, completely synchronized, phase-flipped and AD states respectively. The black lines enclose the region of multi-stability between the completely synchronized and phase-flipped states.}
\end{center}
\end{figure}

The parameter plane of constant coupling delay ($\tau_0$) versus coupling strength ($\epsilon$) with 2 nodes is shown in figure \ref{fig:MG_eq_net_modu_2_eps_tau0_Chaotic}. For each set of values of ($\epsilon$, $\tau_0$) we iterate the system for 300 cycles with a time step of 0.01 and identify the type of dynamics exhibited by the system. The set of initial conditions for time $\in (-\tau_s,\ 0)$ were randomly chosen between 0 and 1. The blue, red, green and yellow colored regions in figure \ref{fig:MG_eq_net_modu_2_eps_tau0_Chaotic} represent unsynchronized, completely synchronized, phase-flipped and AD states respectively. The black circles represent the boundary of the AD region obtained from stability analysis. The white line represents the boundary of the AD region obtained by computing the eigenvalue spectrum of the coupled system using the bifurcation package DDE-BIFTOOL \cite{DDE-BIFTOOL}. As evident from figure \ref{fig:MG_eq_net_modu_2_eps_tau0_Chaotic}, the results via all these methods are found to agree well. The parameter plane in figure \ref{fig:MG_eq_net_modu_2_eps_tau0_Chaotic} has been generated for one set of initial conditions and the phase-flip bifurcation shown corresponds to that set of initial conditions. However for a different set of initial conditions, we observe that the line indicating phase-flip bifurcation  can shift upwards/downwards implying coexistence  of completely synchronized and phase-flipped states. This region of multi-stability is estimated from stability analysis  by computing the LLEs and indicated  as the region enclosed by black lines.
%%%%%%%%%%%%%%%%%%%%
%%%%%%%%%%%%%%%%%%%%

%----------------------------------------------------------------------------------------
%	NATURE OF TRANSITIONS
%----------------------------------------------------------------------------------------

\section{Nature of transitions} \label{sec:NT}

In this section, we study the nature and mechanism of transitions between the different states of emergent dynamics reported in section \ref{sec:ED}.

\begin{figure} [!htbp]
\begin{center}
\includegraphics[scale=0.2]{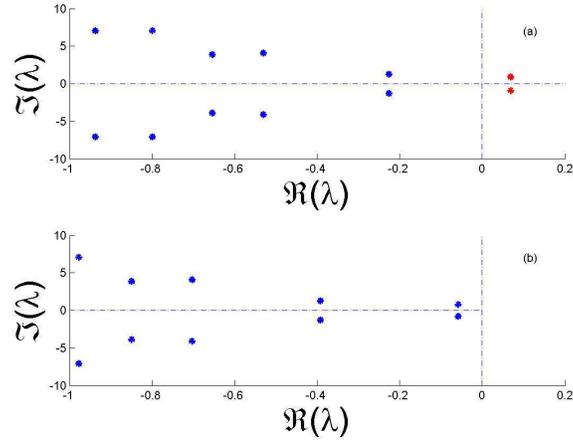}
\caption{\label{fig:MG_eq_net_modu_ES_2_Chaotic}(Colour online) Eigenvalue spectrum of the coupled system at $\tau_0 = 0.2$ and (a) $\epsilon = 0.85$ (i.e., before AD) (b) $\epsilon = 0.9$ (i.e., after AD). With increase in $\epsilon$, the real part of the rightmost eigenvalue crosses the imaginary axis from right to left confirming the existence of AD and route to AD as an inverse Hopf bifurcation.}
\end{center}
\end{figure}

We investigate the route to AD from completely synchronized and phase-flipped states. Figure \ref{fig:MG_eq_net_modu_ES_2_Chaotic} shows the eigenvalue spectrum of the coupled systems at $\epsilon = 0.85$, $\tau_0 = 0.2$ (i.e., before AD) and  $\epsilon = 0.9$, $\tau_0 = 0.2$ (i.e., after AD) respectively. For these parameter values, the system undergoes a transition from the completely synchronized state to AD. With increase in $\epsilon$, the real part of the rightmost eigenvalue crosses the imaginary axis from right to left confirming the existence of AD and the route to AD as an inverse Hopf bifurcation as expected analytically from (\ref{eq:Sign_dl_de}). Qualitatively similar results are obtained in the case of transition from phase-flipped state to AD and hence are not presented here.
%%%%%%%%%%%%%%%%%%%%
%%%%%%%%%%%%%%%%%%%%

\begin{figure} [!htbp]
\begin{center}
\includegraphics[scale=0.25]{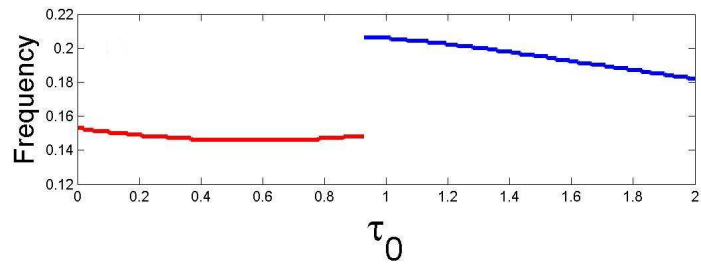}
\caption{\label{fig:MG_eq_net_modu_FFT_tau0_fq_Chaotic}(Colour online) Frequency versus coupling delay ($\tau_0$) at $\epsilon = 0.6$. The red and blue lines represent completely synchronized and phase-flipped states respectively where the discontinuous change in the frequency is clearly evident thus confirming phase-flip bifurcation.} 
\end{center}
\end{figure}

We observe a phase-flip bifurcation in figure \ref{fig:MG_eq_net_modu_2_eps_tau0_Chaotic} as the coupling delay ($\tau_0$) and coupling strength ($\epsilon$) are varied. For $N = 2$ systems, the systems in complete synchrony undergo a phase-flip to attain an anti-phase synchronized state ($\psi \sim \pi$). Phase-flip is accompanied by an abrupt change in the frequency of the synchronized systems \cite{PF_Prasad}. The oscillation frequency of the systems is shown as a function of the coupling delay ($\tau_0$) in figure \ref{fig:MG_eq_net_modu_FFT_tau0_fq_Chaotic} at $\epsilon = 0.6$. The red and blue lines represent completely synchronized and phase-flipped states respectively where the discontinuous change in the frequency is clearly evident thus confirming phase-flip bifurcation.
%%%%%%%%%%%%%%%%%%%%
%%%%%%%%%%%%%%%%%%%%

\begin{figure} [!htbp]
\begin{center}
\includegraphics[scale=0.2]{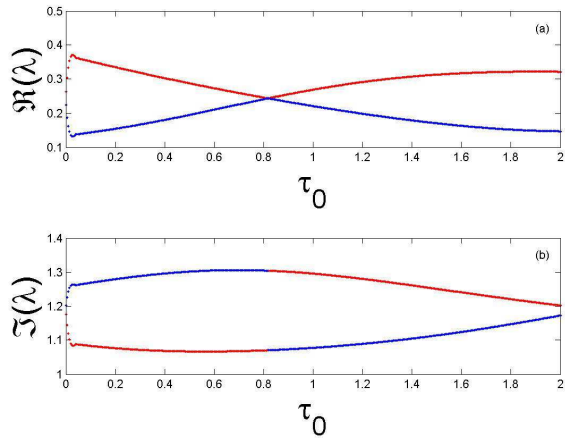}
\caption{\label{fig:MG_eq_net_modu_2_tau0_PFA_Re_Im}(Colour online) Real and imaginary parts of the rightmost and second rightmost eigenvalues are shown in red and blue colors respectively at $\epsilon = 0.6$. (a) Real part of the two rightmost eigenvalues cross each other at $\tau_0$ = 0.818. (b) Interchange of the imaginary part of the two rightmost eigenvalues at $\tau_0$ = 0.818.}
\end{center}
\end{figure}

Further, the nature of phase-flip bifurcation is numerically investigated for greater insight. As reported in \cite{PFA}, for two coupled Landau-Stuart limit cycle oscillators the phase-flip transition is accompanied by an interchange of the imaginary parts of complex pairs of eigenvalues at an avoided crossing of Lyapunov exponents when a parameter is modulated. An investigation of the eigenvalue spectrum of our model in the case of coupling delay ($\tau_0$) reveals that the transition is associated with an interchange of imaginary parts of the rightmost eigenvalues at a value of $\tau_0$ where the real parts of the eigenvalues cross each other as shown in figure \ref{fig:MG_eq_net_modu_2_tau0_PFA_Re_Im}.
%%%%%%%%%%%%%%%%%%%%
%%%%%%%%%%%%%%%%%%%%

\begin{figure} [!htbp]
\begin{center}
\includegraphics[scale=0.15]{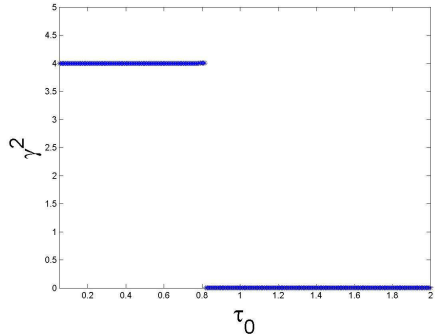}
\caption{\label{fig:MG_eq_net_modu_2_FPA_order_parameter}(Colour online) Order parameter ($\gamma^2$) versus coupling delay ($\tau_0$) at $\epsilon = 0.6$. Prior to the transition, the order parameter has a constant non-zero value while it vanishes identically after the transition implying that the leading eigenvector in the phase-flipped state becomes orthogonal to the leading eigenvector in the completely synchronized state.}
\end{center}
\end{figure}

We construct an order parameter using the eigenvectors corresponding to the leading eigenvalues of the coupled system similar to that considered in \cite{PFA} in order to effectively capture the transition. This is defined as the inner product given by
\begin{equation}
\gamma = <e^{i}(\tau_0') \vert e^{i}(\tau_0)>
\end{equation}
where $e^i(\tau_0)$ is an eigenvector associated with the $i$\textsuperscript{th} rightmost eigenvalue at a coupling delay of $\tau_0$. $\tau_0'$ is a value of the coupling delay on one side of the transition. We consider the value of $\tau_0 = 0.2$ i.e., in the region of complete synchrony and take $\gamma^2$ as the order parameter which detects the transition. Figure \ref{fig:MG_eq_net_modu_2_FPA_order_parameter} shows the variation in the values of the order parameter as a function of $\tau_0$. Prior to the transition, the order parameter has a constant non-zero value while it vanishes identically after the transition implying that the leading eigenvector in the phase-flipped state becomes orthogonal to the leading eigenvector in the completely synchronized state. This can be understood easily by considering the fact that the leading eigenvectors in the completely synchronized and phase-flipped states would be
$\left(\begin{array}{c}
1\\
1
\end{array}\right)\ \textrm{and}\ \left(\begin{array}{c}
1\\
-1
\end{array}\right)$
respectively since the synchronization manifolds in the completely synchronized and phase-flipped states are $x_1 = x_2$ and $x_1 = -x_2$ respectively. Evidently, the eigenvectors are orthogonal to each other.
%%%%%%%%%%%%%%%%%%%%
%%%%%%%%%%%%%%%%%%%%

%----------------------------------------------------------------------------------------
%	PHASE LOCKED CLUSTERS
%----------------------------------------------------------------------------------------

\section{Phase locked clusters} \label{sec:PLC}

So far, we have presented the study for the simplest case of 2 systems. In this section, we explore the dynamics offered by our model in (\ref{eq:Model}) for an arbitrarily large number of systems. We find the network considered is host to richer dynamical behaviour for a relatively large number of systems. We consider the specific cases of $N = 15$ and $N = 17$ i.e., non-prime and prime number of systems respectively with coupling delay.

\begin{figure} [!htbp]
\begin{center}
\includegraphics[width=9cm,height=4cm]{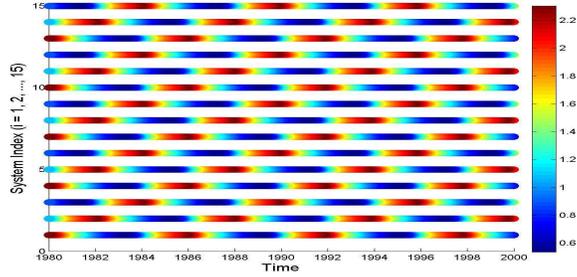}
\caption{\label{fig:MG_eq_net_modu_cluster_N_Chaotic_1}(Colour online) Spatio-temporal plot for $N = 15$ systems at $\epsilon = 0.6$ and $\tau_0 = 1.5$ with color coded values of the dynamical variable ($x_i$). Three phase locked clusters of 5 systems each are formed where the systems exhibit bunching effects by clustering into groups where the systems belonging to a particular group are in complete synchrony with each other.}
\end{center}
\end{figure}

For $N = 15$ systems, in addition to the emergent dynamical states of complete synchrony, phase-flip and AD, we observe that for certain parameter values the systems show a tendency to spread out in phase occupying the entire circle of $2\pi$ but in phase locked clusters. Interestingly, the systems exhibit bunching effects by clustering into groups where the systems belonging to a particular group are in complete synchrony with each other. Depending upon the parameter values, the systems split into 3 clusters of 5 systems each (shown in figure \ref{fig:MG_eq_net_modu_cluster_N_Chaotic_1}) or 5 clusters of 3 systems each. In general, we find that $N$ systems cluster into $N_c$ groups of $m$ systems each where $N = N_c \times m$.
%%%%%%%%%%%%%%%%%%%%
%%%%%%%%%%%%%%%%%%%%

\begin{figure} [!htbp]
\begin{center}
\includegraphics[width=9cm,height=4cm]{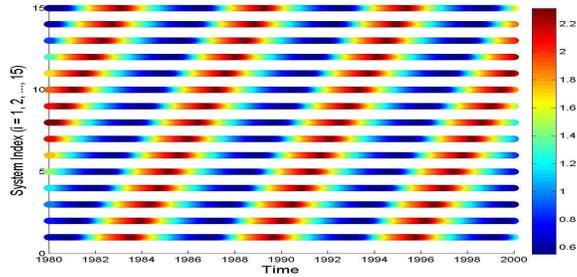}
\caption{\label{fig:MG_eq_net_modu_cluster_N_Chaotic_2}(Colour online) Spatio-temporal plot for $N = 15$ systems at $\epsilon = 0.6$ and $\tau_0 = 0.1$ with color coded values of the dynamical variable ($x_i$). We observe splay states in which the systems split up sequentially in phase.}
\end{center}
\end{figure}

\begin{figure} [!htbp]
\begin{center}
\includegraphics[width=9cm,height=4cm]{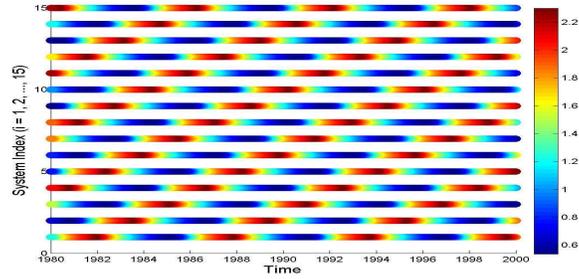}
\caption{\label{fig:MG_eq_net_modu_cluster_N_Chaotic_3}(Colour online) Spatio-temporal plot for $N = 15$ systems at $\epsilon = 0.6$ and $\tau_0 = 1.1$ with color coded values of the dynamical variable ($x_i$). We observe mixed splay states in which the systems arbitrarily distribute out in phase such that the phase difference between two particular systems is $\frac{2\pi}{15}$ but the systems are not arranged sequentially from $i = 1, 2, ..., 15$ unlike the splay state configuration.}
\end{center}
\end{figure}

Corresponding to the phase-flipped state, in this case the systems exhibit `splay states' by splitting up sequentially such that the phase difference between the $i$\textsuperscript{th} and $(i+1)$\textsuperscript{th} system (where $i = 1, 2, ..., 14$) is $\frac{2\pi}{15}$ as shown in figure \ref{fig:MG_eq_net_modu_cluster_N_Chaotic_2} \cite{Splay_State}. However, for certain parameter values, we find that the systems exhibit ``mixed'' splay states by arbitrarily distributing out in phase as shown in figure \ref{fig:MG_eq_net_modu_cluster_N_Chaotic_3} such that the phase difference between two particular systems is $\frac{2\pi}{15}$ but the systems are not arranged sequentially from $i = 1, 2, ..., 15$ unlike the splay state configuration.
%%%%%%%%%%%%%%%%%%%%
%%%%%%%%%%%%%%%%%%%%

%%%%%%%%%%%%%%%%%%%%
%%%%%%%%%%%%%%%%%%%%

For a prime number ($N_p$) of systems, we can have only $N_c = 1$ ($m = N_p$)  or $N_c = N_p$ ($m = 1$) implying that the systems are either in-phase or distributed in phase. Therefore, we find the absence of such clustering effects in our network for a prime number ($N_p$) of systems. In such cases, the systems exhibit splay states or ``mixed'' splay states with a phase difference of $\frac{2\pi}{N_p}$ between successive systems i.e., $\frac{2\pi}{17}$ for $N_p = 17$ systems.

\begin{figure} [!htbp]
\begin{center}
\includegraphics[scale=0.2]{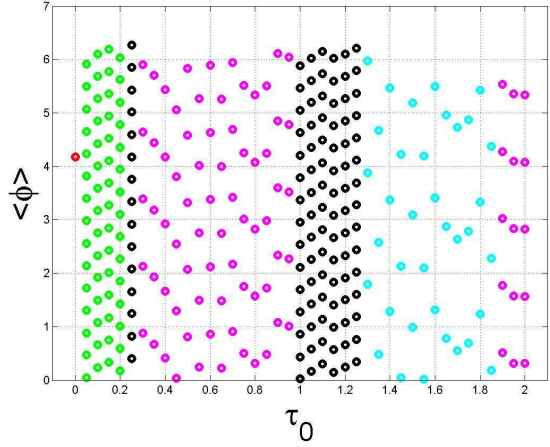}
\caption{\label{fig:Average_Phase}(Colour online) Average phase of $N = 15$ systems as a function of the coupling delay ($\tau_0$) at $\epsilon = 0.6$ showing the transition between the various phase locked states. The red, green and black colored regions represent completely synchronized, splay and mixed splay states respectively. The phase locked clusters for 15 systems are represented by cyan and magenta colored regions where the systems split into 3 clusters of 5 systems each and into 5 clusters of 3 systems each respectively.
}
\end{center}
\end{figure}

For 15 systems we find that increasing $\tau_0$ at a fixed value of $\epsilon = 0.6$ enables the systems to undergo a transition from complete synchrony to splay states to ``mixed'' splay states to 5 cluster state to ``mixed'' splay states to 3 cluster state to 5 cluster state. In general, for a non-prime number of systems we find such alternate regions of phase locked clusters separated by regions of ``mixed'' splay states. This is clearly illustrated for the case of 15 systems in figure \ref{fig:Average_Phase} where the average phase ($<\phi>$) is calculated as a function of the coupling delay ($\tau_0$) at $\epsilon = 0.6$. The blue, red, green, black and yellow colored regions represent unsynchronized, completely synchronized, splay, mixed splay and AD states respectively. For 17 systems, we observe a greater tendency to form ``mixed'' splay states compared to that for the splay states. Increasing the number of systems $N$, we obtain qualitatively similar results for a non-prime and prime number of systems as described for 15 and 17 systems respectively.
%%%%%%%%%%%%%%%%%%%%
%%%%%%%%%%%%%%%%%%%%

%----------------------------------------------------------------------------------------
%	EMERGENT DYNAMICS IN THE MODEL WITH DIGITAL COUPLING
%----------------------------------------------------------------------------------------

\section{Emergent dynamics with digital coupling} \label{sec:Model_Hybrid}

Previously, we considered the continuous flow of data from one node to another in our model (in figure \ref{fig:Model}). However, in a digital communication system, this is unlikely and data sent from one component is both sampled and quantized (digitized) before it is received by another which forms the subject of the present section.

We consider a sampling scheme where the value of a dynamical variable at an integer multiple of sampling time is used for coupling during the entire subsequent interval of sampling time. Then, the dynamical equation of the network (in (\ref{eq:Model})) with the digitization of coupling becomes:
\begin{equation} \label{eq:Model_DC_IVS}
\dot{x}_i = af(x_{i\tau_s})-bx_i + \epsilon \left[ \frac{x_j(\left[ \frac{t}{ts}\right] ts-\tau_0)}{qstep}\right] qstep
\end{equation}
for $i = 1, 2, 3, \ldots, N-1, N\ \&\ j = N, 1, 2, \ldots, N-1$ respectively . $[.]$ stands for the greatest integer function, $ts$ is the sampling time, $qstep$ is the step size in quantizing the dynamical variable $x_i$.

\begin{figure} [!htbp]
\begin{center}
\includegraphics[scale=0.2]{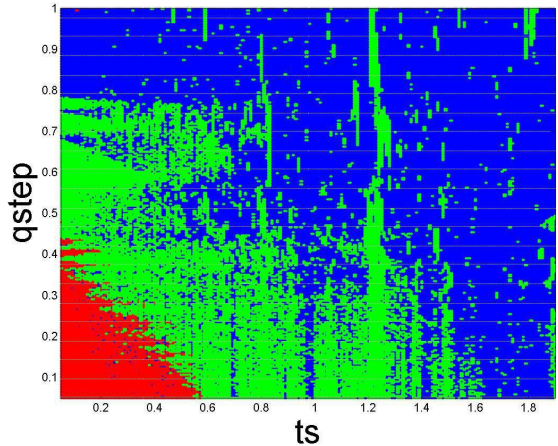}
\caption{\label{fig:MG_eq_net_modu_2_digital_coupling_ts_qstep} (Colour online) Parameter plane of $qstep$ versus $ts$ for 2 systems at $\epsilon = 0.6$ and $\tau_0 = 0.4$. The blue, red and green colored regions represent unsynchronized, completely synchronized and phase-flipped states respectively.}
\end{center}
\end{figure}

For $\epsilon = 0.6$ and $\tau_0 = 0.4$, we explore the parameter plane of $qstep$ versus $ts$ for 2 systems in figure \ref{fig:MG_eq_net_modu_2_digital_coupling_ts_qstep} and demonstrate the occurrence of phase-flip in the system due to sampling and quantization. This can be understood as sampling time creating an additional delay in the system since it takes a value from the past and uses it during the subsequent sampling interval. We note that so far phase-flip due to coupling delay alone has been reported.
%%%%%%%%%%%%%%%%%%%%
%%%%%%%%%%%%%%%%%%%%

\begin{figure} [!htbp]
\begin{center}
\includegraphics[scale=0.2]{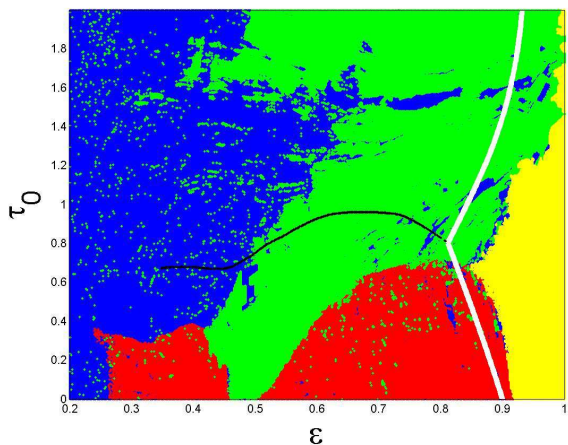}
\caption{\label{fig:MG_eq_net_modu_2_digital_coupling_eps_tau0} (Colour online) Parameter plane of $\tau_0$ versus $\epsilon$ for 2 systems at $ts = 0.35$ and $qstep = 0.2$. We find an overall increase of unsynchronized dynamics (in blue) while reduction in particularly complete synchronization (in red). We also find that the systems undergo phase-flip (in green) at a much lower value of coupling delay as opposed to the analog based coupling situation considered earlier in figure \ref{fig:MG_eq_net_modu_2_eps_tau0_Chaotic}. We also observe AD (in yellow) but now the region has reduced in size.}
\end{center}
\end{figure}

Further, to explore the effect of digitization for all values of $\epsilon$ and $\tau_0$, we study the possible emergent states in the $(\epsilon$, $\tau_0)$ parameter plane at $ts = 0.35$ and $qstep = 0.2$ (figure \ref{fig:MG_eq_net_modu_2_digital_coupling_eps_tau0}). Then, we compare it with the results obtained for analog coupling illustrated in figure \ref{fig:MG_eq_net_modu_2_eps_tau0_Chaotic}. The color code for the emergent states in figures \ref{fig:MG_eq_net_modu_2_eps_tau0_Chaotic} and \ref{fig:MG_eq_net_modu_2_digital_coupling_eps_tau0} are the same. The black line in figure \ref{fig:MG_eq_net_modu_2_digital_coupling_eps_tau0} represents the parameter values at which the phase-flip transition occurs in the analog coupling setting and the white line shows the transition to AD (figure \ref{fig:MG_eq_net_modu_2_eps_tau0_Chaotic}).

In the hybrid case where sampling and quantization are both present, we find reduced regions of synchronization particularly complete synchronization. For a given value of coupling strength, we generally find that the systems undergo phase-flip at a much lower value of coupling delay as opposed to the analog based coupling situation considered earlier. This shows that digitization lowers the value of coupling delay at which phase-flip transition occurs. Thus, the effect of digitization can be considered as equivalent to increased coupling delay. Also, as is clear from figure \ref{fig:MG_eq_net_modu_2_digital_coupling_eps_tau0}, the region of AD is reduced due to digitized coupling where the transition curve of AD is pushed to higher values of $\epsilon$.
%%%%%%%%%%%%%%%%%%%%
%%%%%%%%%%%%%%%%%%%%

%----------------------------------------------------------------------------------------
%	CONCLUSION
%----------------------------------------------------------------------------------------

\section{Conclusion} \label{sec:Conclusion}

We report the rich variety of dynamical states like amplitude death, isochronal synchrony, phase-flip, splay states and phase locked cluster states in a ring network of delay dynamical systems connected via unidirectional positive feedback with constant delay in coupling. Our analysis is carried out with a prototypical time delay system, viz. Mackey-Glass system. The role of delayed connections in inducing various states like AD and splay states have been reported earlier in similar context but with diffusive coupling. Our work indicates that positive feedback can give rise to similar interesting dynamical states and also control and stabilise the systems to steady states. Delay coupled delay dynamical systems generally exhibit lag/anticipatory synchronization. We observe isochronal synchrony with control for systems coupled with unidirectional positive feedback with delay.

In the specific context of 2 MG systems on the network, the linear stability analysis is used to isolate the region in the parameter space where the coupled system exhibits AD. Further, the eigenvalue spectrum of the system has been computed to determine the exact parameter values for AD. These results have been verified by direct numerical simulation of our model.

The study presented also includes detailed analysis of the transitions between various different dynamics. Thus, the eigenvalue spectrum of the coupled system reveals that it goes to AD via Hopf bifurcation. The phase-flip bifurcation is marked by a discrete change in the frequency of the coupled systems and is associated with an interchange of the imaginary parts of the eigenvalues at a crossing of the real part of eigenvalues. The order parameter constructed following \cite{PFA} captured this transition effectively. For certain parameter values, we capture multi-stability between the synchronized states in our system (upon changing the initial conditions).

Our study on the ring network with larger number of systems shows that in addition to the emergent dynamics of complete synchrony, phase-flip and AD, the systems exhibit bunching effects by clustering into groups such that the systems belonging to a particular group are in complete synchrony. They also exhibit different phase behaviour like splay states and  mixed splay states for certain values of coupling strength and delay. We identify regions in this parameter plane corresponding to various possible dynamical states using detailed numerical analysis. We note that isochronal synchronization and splay states have been recently experimentally observed in optoelectronic periodic oscillators coupled in ring topology \cite{Isochronal_Splay_Experiment} and also in pulse coupled neuronal networks \cite{PF_Universal}.

Further, we include the effect of digital coupling which is relevant in a practical implementation of such network structures. We find that digitization can induce phase-flip bifurcation which facilitates anti-synchronization at lower values of coupling delay. This also reduces regions of AD.

The results reported are for identical MG systems forming the nodes of the network with the individual MG systems in the chaotic regime. However, we obtain qualitatively similar results for individual MG systems in the hyperchaotic regime as well as for non-identical MG systems (with slight parameter mismatches) with analog coupling in the same dynamical configuration.

\section*{References}

\bibliography{References}

\end{document}